\documentclass[english, final, 5p, times, sort&compress]{iopart}
\usepackage{mathptmx}
\usepackage[latin9]{inputenc}
\usepackage{geometry}
\usepackage[dvipsnames]{xcolor}
\usepackage{babel}
\usepackage{array}
\usepackage{textcomp}
\usepackage{multirow}
\usepackage{soul}
\usepackage{iopams}
\usepackage{amsthm}
\usepackage{amssymb}
\usepackage{graphicx}
\usepackage[unicode=true,
 bookmarks=true,bookmarksnumbered=false,bookmarksopen=false,
 breaklinks=false,pdfborder={0 0 0},backref=false,colorlinks=true]
 {hyperref}
\hypersetup{
 pdfauthor={Th. J.-Y. Derrien},
 citecolor=blue, urlcolor=blue, linkcolor=blue}

\makeatletter

\providecommand{\tabularnewline}{\\}

\newcommand{\revadd}[1]{#1}

\@ifundefined{showcaptionsetup}{}{%
 \PassOptionsToPackage{caption=false}{subfig}}
\usepackage{subfig}
\makeatother

\begin{document}

\title[Properties of SPPs on lossy materials]{Properties of Surface Plasmon Polaritons on lossy materials: Lifetimes, periods and excitation conditions}

\selectlanguage{english}%

\author{Thibault J.-Y. Derrien}
\ead{derrien@fzu.cz}

\address{HiLASE Centre, Institute of Physics, Czech Academy of Sciences, \\
Za Radnic\'{i} 828, 25241 Doln\'{i} B\v{r}e\v{z}any, Czech Republic}
\address{Bundesanstalt f\"{u}r Materialforschung und \textendash pr\"{u}fung (BAM), \\
Unter den Eichen 87, D-12205 Berlin, Germany}

\author{J\"{o}rg Kr\"{u}ger}
\address{Bundesanstalt f\"{u}r Materialforschung und \textendash pr\"{u}fung (BAM), \\
Unter den Eichen 87, D-12205 Berlin, Germany}

\author{J\"{o}rn Bonse}
\address{Bundesanstalt f\"{u}r Materialforschung und \textendash pr\"{u}fung (BAM), \\
Unter den Eichen 87, D-12205 Berlin, Germany}

\begin{indented}
\item[]{May 21st, 2016}
\end{indented}

\begin{abstract}
The possibility to excite Surface Plasmon Polaritons (SPPs) at the interface between two media depends on the optical properties of both media
and geometrical aspects. Specific conditions allowing the coupling
of light with a plasmon-active interface must be satisfied. Plasmonic
effects are well described in noble metals where the imaginary part of
the dielectric permittivity is often neglected (\textquotedblleft perfect
medium approximation\textquotedblright ). However, some systems exist
for which such approximation cannot be applied, hence requiring a
refinement of the common SPP theory. In this context, several properties
of SPPs such as excitation conditions, period of the electromagnetic
field modulation and SPP lifetime then may strongly
deviate from that of the perfect medium approximation. In this paper, calculations
taking into account the imaginary part of the dielectric permittivities
are presented. The model identifies analytical terms which should
not be neglected in the mathematical description of SPPs on lossy materials. These calculations
are applied to numerous material combinations resulting in a prediction of the 
corresponding SPP features. A list of plasmon-active interfaces
is provided along with a quantification of the above mentioned SPP properties 
in the regime where the perfect medium approximation is not applicable. 
\end{abstract}

\vspace{2pc}
\noindent{\it Keywords}: surface plasmon polaritons, lossy materials, plasmon lifetime

\submitto{\JOPT : The final version is available at the following address: http://dx.doi.org/10.1088/2040-8978/18/11/115007}
% \linenumbers

\maketitle

% \tableofcontents

\section{Introduction}

%\paragraph*{Fundamental plasmonics and applications}

Surface Plasmon Polaritons (SPPs) are collective oscillations of electrons	
occuring at the interface of materials. More than hundred years after their discovery \cite{Maier2007}, SPPs have promoted new applications in many fields such as microelectronics \cite{MacDonald2009}, photovoltaics \cite{Atwater2010}, near-field sensing \cite{Otto1974,Bell1975a,Kabashin2009}, laser techonology \cite{Berini2011,Park2011}, photonics \cite{Barnes2003,Liu2008a}, meta-materials design \cite{Shalaev2007}, high order harmonics generation \cite{Kim2008b}, or charged particles acceleration \cite{Genevet2010, Purvis2013}.

Most of these applications are based on expensive noble metals
such as gold, silver or platinum, as these materials greatly support the plasmonic phenomena, exhibit 
very small (plasmonic) losses and the experimental results match well with the associated theory \cite{Maier2007, Otto1974, Berini2011, Pettit1975,Raether1986}. 
\revadd{Although there were numerous studies addressing SPPs in lossy materials \cite{Bell1973,Alexander1974a,Ward1974,Ward1975a,Kovener1976,Kovener1977,Ritchie1977, Laglois1978,Reinisch1979,Boardman1984, Norrman2013,Lee2013a,Martinez-Herrero2015}, some specific aspects remain to be investigated. }

In this paper, a mathematical condition
for SPP excitation at flat interfaces is provided. This approach includes the widely accepted theory but reveals
a wider (material dependent) domain of SPP excitation than predicted by the existing literature. The importance of the terms originating from losses is underlined and complemented by formula of the SPP near-field period and lifetime.

\section{Excitation conditions for Surface Plasmon Polariton in lossy materials}

\begin{figure}
\begin{centering}
\includegraphics[width=8cm]{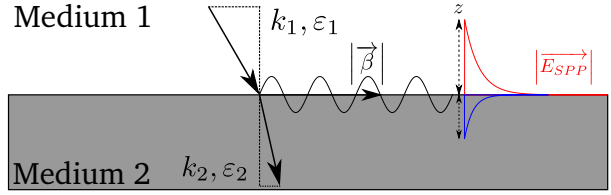}
\label{fig:Fig1}
\par\end{centering}

\caption{\label{fig:Scheme}Scheme of a planar interface between medium 1
and 2 at which Surface Plasmon Polaritons are considered. The wavevector components along the surface normal axis are indicated as $k_{1,2}$ and $\varepsilon_{1,2}$
are the complex dielectric permittivities of both media. $\beta$ is the complex wavenumber of the SPPs propagating along the surface plane. }
\end{figure}

At a planar interface between two different materials, the electric field components ($E_{x,y,z}$) and magnetic field components ($H_{x,y,z}$) can be calculated by solving the Helmholtz equation for Transverse Magnetic (TM) and Transverse Electric (TE) boundary conditions \cite{Maier2007,Raether1986,Zayats2005}. For the geometry provided in Fig. \ref{fig:Scheme}, the mathematical solutions crucially depend on two complex-valued properties: the dielectric permittivities $\varepsilon_{i}\in\mathbb{C}$ (linked to the optical refractive indices $n_{i}$ of medium $i$ by $\varepsilon_{i}=n_{i}^{2}$), and the complex-valued wavenumbers $k_{i}\in\mathbb{C}$ (associated with electromagnetic field modes in the medium $i$). At the interface between two media (1 and 2), the conservation of light momentum results in the condition \cite{Maier2007} 
\begin{equation}
k_{1,2}^{2}=\beta^{2}-k_{0}^{2}\varepsilon_{1,2}, \label{eq:DefinitionOfSPPwaveNumber}
\end{equation}where $\beta\in\mathbb{C}$ is the SPP wavenumber along the interface, $k_{0}=\frac{\omega}{c}$ is the wavenumber of the incident light \revadd{($\omega \in \mathbb{R}$}: light angular frequency, $c$: light velocity in vacuum). 

In TM geometry, the continuity conditions for the electromagnetic fields results in the relation \cite{Maier2007,Raether1986,Zayats2005} 
\begin{equation}
\frac{k_{1}}{\varepsilon_{1}}+\frac{k_{2}}{\varepsilon_{2}}=0.\label{eq:MainSPPcondition}
\end{equation}Equation (\ref{eq:MainSPPcondition}) represents the \emph{dispersion relation of Surface Plasmon Polaritons} at the interface between two semi-infinite media. \revadd{The signs of $\Re e(k_1)$ and $\Re e(k_2)$ were taken positive here, accounting for the exponential decay of the electromagnetic field amplitude in the direction perpendicular to (away from) the interface.}
The combination of Eqs. (\ref{eq:DefinitionOfSPPwaveNumber}) and (\ref{eq:MainSPPcondition}) provides \revadd{the solutions} of the SPP wavenumber
$\beta$: \revadd{
\begin{eqnarray}
\beta & = & \pm \frac{\omega}{c}\sqrt{\frac{\varepsilon_{1}\varepsilon_{2}}{\varepsilon_{1}+\varepsilon_{2}}}.\label{eq:BetaDefinition}
\end{eqnarray}
}

\revadd{In order to address the mathematical description of damped surface electromagnetic waves, in the following, we selected the positive branch of $\beta$.}

\revadd{It must noted that in a generalized view, the characterization of lossy waves can be treated by calculating an observable response function  $F(\omega, \beta)$ which allows to construct a dispersion relation by locating its complex zeros/poles \cite{ Kovener1976, Ritchie1977}. }

\revadd{As already noted by Ritchie et al. \cite{Ritchie1977}, when damping is relevant, the dispersion relation $\omega (\beta)$ for $\beta \in \mathbb{R}$ may have complex solutions ($\omega \in \mathbb{C}$). Conversely, if $\omega$ is real-valued, $\beta$ may be complex-valued. Although straightforward in a theoretical framework, there is some ambiguity about the significance of complex values of $\omega$ or $\beta$ in the interpretation of experiments \cite{Ritchie1977}. }

\revadd{In experiments it may be difficult to observe temporal or spatial decay
of a resonance due to its rapidity or smallness. The properties
of such excitations are usually extracted from the transfer of energy and momentum 
to the system, involving both real $\omega$ and real $k$. 
As an example, dispersion relations have been  determined by Attenuated Total Reflection  (ATR) in Otto configuration \cite{Otto1974}. In this approach, a beam is totally reflected at the basis plane of an optical prism. Excitation of SPP  in a neighbored metal surface may be realized via coupling through a gap of a dielectric medium (air). SPP manifest as drops in the totally reflected signal, when momentum matching between light and SPP occurs. Experimentally, as underlined by Kovener et al. \cite{Kovener1976}, this can be realized either for a variation of frequency $\omega$ at a fixed angle of incidence $\theta$, or via a variation of $\theta$ at  a fixed $\omega$.  The first procedure produces dispersion curves with a specific "bend back" feature, while the second procedure results in curves without that feature \cite{Kovener1976}. }

To calculate the excitation conditions of SPPs at the interface between two arbitrary (absorbing) media, a mathematical analysis is presented in the following. 

\subsection{Sign analysis on the dispersion relation\label{sec:Conditions-for-SPP}}

SPPs can be excited only if the dispersion relation [Eq. (\ref{eq:MainSPPcondition})] is fulfilled. In order to extract the SPP excitation conditions from the dispersion relation, a sign analysis can be performed on the real and imaginary parts of Eq. (\ref{eq:MainSPPcondition}), which can be mathematically developed to: 
\begin{equation}
\label{eq:ExactLin1}
\cases{
\Re e\left(\frac{k_{1}}{\varepsilon_{1}}+\frac{k_{2}}{\varepsilon_{2}}\right) & = 0\\
\Im m\left(\frac{k_{1}}{\varepsilon_{1}}+\frac{k_{2}}{\varepsilon_{2}}\right) & = 0}
\end{equation}
By assuming \revadd{\[\Re e\left(k_{1}\right) > 0 \mbox{ and } \Re e\left(k_{2}\right)>0, \]} this equation can be used to deduce a constraint on the sign of real part of the dielectric permittivities $\varepsilon_{1,2}$, resulting in
\begin{equation}
\Re e\left(\frac{\varepsilon_{1}}{\varepsilon_{2}}\right)<0.\label{eq:ConditionSymmetricSPP}
\end{equation}
Equation (\ref{eq:ConditionSymmetricSPP}) defines the first necessary condition for excitation of Surface Plasmon Polaritons, which is equivalent to
\begin{equation}
\Re e\left(\varepsilon_{1}\right)\times\Re e\left(\varepsilon_{2}\right)+\underbrace{\Im m\left(\varepsilon_{1}\right)\times\Im m\left(\varepsilon_{2}\right)}_{\mbox{important for lossy materials}}<0.\label{eq:ConditionsSPP1}
\end{equation}
The physical meaning of Eq. (\ref{eq:ConditionsSPP1}), named \emph{Condition 1} for SPP excitation, is the following: in presence of a perfect dielectric medium (e.g. for $\Im m\left(\varepsilon_{1}\right)=0$ or $\Im m\left(\varepsilon_{2}\right)=0$), Eq. (\ref{eq:ConditionsSPP1}) implies that SPPs can only be excited at a dielectric - metal interface. This is usually fulfilled when $\Im m\left(\varepsilon_{1,2}\right)\ll\left|\Re e\left(\varepsilon_{1,2}\right)\right|$. Then Eq. (\ref{eq:ConditionsSPP1}) reduces to the widely accepted expression \cite{Maier2007,Zayats2005,Bonch-Bruevich1992}
\begin{equation}
\Re e\left(\varepsilon_{1}\right)\times\Re e\left(\varepsilon_{2}\right)<0.\label{eq:Condition1-Re}
\end{equation}

However, in presence of \emph{two absorbing media }(having then $\Im m\left(\varepsilon_{1}\right)>0$ and $\Im m\left(\varepsilon_{2}\right)>0$), the physical meaning is less intuitive. The condition given by Eq. (\ref{eq:ConditionsSPP1}) is more complex due to non-vanishing contributions of the imaginary parts of the dielectric
permittivities. Consequences of this additional term, important for lossy materials, will be discussed in Section \ref{sec:exploration}. 

\subsection{Perfect medium approximation and beyond}

Assuming purely real-valued dielectric permittivities, Eq. (\ref{eq:BetaDefinition}) is typically used to derive another condition for SPP excitation.  This approach is called \emph{Perfect Medium Approximation} (PMA) and will be outlined in the following. 
For simplicity, we adopt  the following notations: $\varepsilon_{i}'=\Re e\left(\varepsilon_{i}\right)$
and $\varepsilon_{i}''=\Im m\left(\varepsilon_{i}\right)$. If $\varepsilon''_{1}$ and $\varepsilon''_{2}=0$, then $\beta$ is real-valued ($\mathbb{R}$), 
%or equivalently
and Eq. (\ref{eq:BetaDefinition}) becomes

\begin{eqnarray}
% \begin{cases}
\label{eq:ConditionSPP2}
\cases{\Re e\left(\frac{\varepsilon_{1}\varepsilon_{2}}{\varepsilon_{1}+\varepsilon_{2}}\right)>0\\
\Im m\left(\frac{\varepsilon_{1}\varepsilon_{2}}{\varepsilon_{1}+\varepsilon_{2}}\right)=0.}
\end{eqnarray}

The latter can be rewritten as
\revadd{\[
\cases{
\frac{\left(\varepsilon_{1}'\varepsilon_{2}'-\varepsilon_{1}''\varepsilon_{2}''\right)\left(\varepsilon_{1}'+\varepsilon_{2}'\right)+\left(\varepsilon_{1}''\varepsilon_{2}'+\varepsilon_{1}'\varepsilon_{2}''\right)\left(\varepsilon_{1}''+\varepsilon_{2}''\right)}{\left(\varepsilon_{1}'+\varepsilon_{2}'\right)^{2}+\left(\varepsilon_{1}''+\varepsilon_{2}''\right)^{2}} & $>$ 0 \\
\frac{\left(\varepsilon_{1}''\varepsilon_{2}'+\varepsilon_{1}'\varepsilon_{2}''\right)\left(\varepsilon_{1}'+\varepsilon_{2}'\right)-\left(\varepsilon_{1}'\varepsilon_{2}'-\varepsilon_{1}''\varepsilon_{2}''\right)\left(\varepsilon_{1}''+\varepsilon_{2}''\right)}{\left(\varepsilon_{1}'+\varepsilon_{2}'\right)^{2}+\left(\varepsilon_{1}''+\varepsilon_{2}''\right)^{2}} & = 0}
%\end{cases}
\]}
which is strictly equivalent to
\begin{equation}
\frac{\varepsilon_{1}'\varepsilon_{2}'}{\varepsilon_{1}'+\varepsilon_{2}'}>0\label{eq:RealCasesCondition2}.
\end{equation}
Condition [Eq. (\ref{eq:RealCasesCondition2})] implies

\begin{equation}
\left| \varepsilon_{2}' \right| > \varepsilon_{1}'\label{eq:RealCondition}
\end{equation}
and widely used in literature \cite{Raether1986, Bonch-Bruevich1992}. In particular, when medium 1 is air ($\varepsilon_1=1$), the joint application of Eqs. (\ref{eq:Condition1-Re}) and (\ref{eq:RealCondition}) leads to the well admitted condition
\begin{equation}
\Re e\left(\varepsilon_2 \right)<-1 \label{eq:BonchCondition}.
\end{equation} 

However, beyond the perfect medium approximation, it must be noted that Eq. (\ref{eq:BetaDefinition}) can be treated using fully complex permittivity values since $\beta$ is defined in $\mathbb{C}$ for \emph{any} value of the dielectric permittivities $\varepsilon_{i}$. As a consequence, in presence of one (or more) "lossy" materials, e.g., when $\Im m\left(\varepsilon_{1}\right)>0$ or $\Im m\left(\varepsilon_{2}\right)>0$, \emph{there is no other restriction for SPP excitation than the Condition 1} given by Eq. (\ref{eq:ConditionsSPP1}). In other words, performing an \emph{ad-hoc} restriction of the dielectric permittivity to its real part(s) may lead to an oversimplified SPP excitation condition. 

\subsection{Exploration of plasmon-active material combinations \label{sec:exploration}}

In this section, the plasmon activity of a wide set of material combinations is explored by comparing predictions of the PMA [real-valued Eqs. (\ref{eq:Condition1-Re}) and (\ref{eq:RealCasesCondition2})] with the more general case [complex-valued Eq. (\ref{eq:ConditionsSPP1})]. For that, if not stated differently, data of the optical constants were taken from Ref. \cite{Palik1985}. Values are listed in \ref{sec:AppendixOptical}. 

In a first step, a selection of different metals exposed to air are analyzed. The results for the SPP-activity for 12 different noble and transition metals are provided in Tab. \ref{tab:SPP-active-interfaces-air}. Here we restrict the study to two wavelengths frequently used in laser processing: 800 nm and 400 nm. However, these calculations can be generalized to other material combinations and wavelengths. The SPP-activity of the metal-air interfaces is indicated by a green tick (\textcolor{ForestGreen}{$\surd$}), whereas the interfaces which do not support SPPs are marked by a red cross (\textcolor{red}{$\times$}). At both wavelengths the generalized model [Eq. (\ref{eq:ConditionsSPP1})] predicts similar results  as the PMA. Interestingly, and opposed to the PMA, Niobium (Nb) is predicted to support SPPs when interfaced with air, upon 400 nm irradiation. 
\begin{table}
\begin{centering}
\begin{tabular}{c|cc|cc}
\multicolumn{1}{c}{} & \multicolumn{2}{c}{} & \multicolumn{2}{c}{}\tabularnewline
\hline 
\hline 
\multirow{2}{*}{{\footnotesize{}Interface}} & \multicolumn{2}{c|}{{\footnotesize{}$\lambda=800$ nm}} & \multicolumn{2}{c}{{\footnotesize{}$\lambda=400$ nm}}\tabularnewline
 & {\footnotesize{}$\Im m\left(\varepsilon_{2}\right)=0$}  & {\footnotesize{}$\Im m\left(\varepsilon_{2}\right)\neq0$}  & {\footnotesize{}$\Im m\left(\varepsilon_{2}\right)=0$}  & {\footnotesize{}$\Im m\left(\varepsilon_{2}\right)\neq0$}\tabularnewline
\hline 
{\footnotesize{}Air/Ag}  & \textbf{\textcolor{ForestGreen}{\footnotesize{}$\surd$}}\textcolor{ForestGreen}{{} } & \textbf{\textcolor{ForestGreen}{\footnotesize{}$\surd$}}\textcolor{ForestGreen}{{} } & \textbf{\textcolor{ForestGreen}{\footnotesize{}$\surd$}}\textcolor{ForestGreen}{{} } & \textbf{\textcolor{ForestGreen}{\footnotesize{}$\surd$}}\tabularnewline
{\footnotesize{}Air/Al}  & \textbf{\textcolor{ForestGreen}{\footnotesize{}$\surd$}}\textcolor{ForestGreen}{{} } & \textbf{\textcolor{ForestGreen}{\footnotesize{}$\surd$}}\textcolor{ForestGreen}{{} } & \textbf{\textcolor{ForestGreen}{\footnotesize{}$\surd$}}\textcolor{ForestGreen}{{} } & \textbf{\textcolor{ForestGreen}{\footnotesize{}$\surd$}}\tabularnewline
{\footnotesize{}Air/Au}  & \textbf{\textcolor{ForestGreen}{\footnotesize{}$\surd$}}\textcolor{ForestGreen}{{} } & \textbf{\textcolor{ForestGreen}{\footnotesize{}$\surd$}}\textcolor{ForestGreen}{{} } & \textbf{\textcolor{ForestGreen}{\footnotesize{}$\surd$}}\textcolor{ForestGreen}{{} } & \textbf{\textcolor{ForestGreen}{\footnotesize{}$\surd$}}\tabularnewline
{\footnotesize{}Air/Cr}  & \textbf{\textcolor{ForestGreen}{\footnotesize{}$\surd$}}\textcolor{ForestGreen}{{} } & \textbf{\textcolor{ForestGreen}{\footnotesize{}$\surd$}}\textcolor{ForestGreen}{{} } & \textbf{\textcolor{ForestGreen}{\footnotesize{}$\surd$}}\textcolor{ForestGreen}{{} } & \textbf{\textcolor{ForestGreen}{\footnotesize{}$\surd$}}\tabularnewline
{\footnotesize{}Air/Cu}  & \textbf{\textcolor{ForestGreen}{\footnotesize{}$\surd$}}\textcolor{ForestGreen}{{} } & \textbf{\textcolor{ForestGreen}{\footnotesize{}$\surd$}}\textcolor{ForestGreen}{{} } & \textbf{\textcolor{ForestGreen}{\footnotesize{}$\surd$}}\textcolor{ForestGreen}{{} } & \textbf{\textcolor{ForestGreen}{\footnotesize{}$\surd$}}\tabularnewline
{\footnotesize{}Air/Fe}  & \textbf{\textcolor{ForestGreen}{\footnotesize{}$\surd$}}\textcolor{ForestGreen}{{} } & \textbf{\textcolor{ForestGreen}{\footnotesize{}$\surd$}}\textcolor{ForestGreen}{{} } & \textbf{\textcolor{ForestGreen}{\footnotesize{}$\surd$}}\textcolor{ForestGreen}{{} } & \textbf{\textcolor{ForestGreen}{\footnotesize{}$\surd$}}\tabularnewline
\textbf{\textcolor{blue}{\footnotesize{}Air/Nb}}\textbf{\textcolor{blue}{{} }} & \textbf{\textcolor{ForestGreen}{\footnotesize{}$\surd$}}\textcolor{ForestGreen}{{} } & \textbf{\textcolor{ForestGreen}{\footnotesize{}$\surd$}}\textcolor{ForestGreen}{{} } & \textbf{\textcolor{red}{\footnotesize{}$\times$}}{\footnotesize{} } & \textbf{\textcolor{ForestGreen}{\footnotesize{}$\surd$}}\tabularnewline
{\footnotesize{}Air/Ni}  & \textbf{\textcolor{ForestGreen}{\footnotesize{}$\surd$}}\textcolor{ForestGreen}{{} } & \textbf{\textcolor{ForestGreen}{\footnotesize{}$\surd$}}\textcolor{ForestGreen}{{} } & \textbf{\textcolor{ForestGreen}{\footnotesize{}$\surd$}}\textcolor{ForestGreen}{{} } & \textbf{\textcolor{ForestGreen}{\footnotesize{}$\surd$}}\tabularnewline
{\footnotesize{}Air/Pt}  & \textbf{\textcolor{ForestGreen}{\footnotesize{}$\surd$}}\textcolor{ForestGreen}{{} } & \textbf{\textcolor{ForestGreen}{\footnotesize{}$\surd$}}\textcolor{ForestGreen}{{} } & \textbf{\textcolor{ForestGreen}{\footnotesize{}$\surd$}}\textcolor{ForestGreen}{{} } & \textbf{\textcolor{ForestGreen}{\footnotesize{}$\surd$}}\tabularnewline
{\footnotesize{}Air/Ti}  & \textbf{\textcolor{ForestGreen}{\footnotesize{}$\surd$}}\textcolor{ForestGreen}{{} } & \textbf{\textcolor{ForestGreen}{\footnotesize{}$\surd$}}\textcolor{ForestGreen}{{} } & \textbf{\textcolor{ForestGreen}{\footnotesize{}$\surd$}}\textcolor{ForestGreen}{{} } & \textbf{\textcolor{ForestGreen}{\footnotesize{}$\surd$}}\tabularnewline
{\footnotesize{}Air/V } & \textbf{\textcolor{red}{\footnotesize{}$\times$}}{\footnotesize{} } & \textbf{\textcolor{red}{\footnotesize{}$\times$}}{\footnotesize{} } & \textbf{\textcolor{ForestGreen}{\footnotesize{}$\surd$}}\textcolor{ForestGreen}{{} } & \textbf{\textcolor{ForestGreen}{\footnotesize{}$\surd$}}\tabularnewline
{\footnotesize{}Air/W}  & \textbf{\textcolor{red}{\footnotesize{}$\times$}}{\footnotesize{} } & \textbf{\textcolor{red}{\footnotesize{}$\times$}}{\footnotesize{} } & \textbf{\textcolor{red}{\footnotesize{}$\times$}}{\footnotesize{} } & \textbf{\textcolor{red}{\footnotesize{}$\times$}}\tabularnewline
\hline 
\hline 
\multicolumn{1}{c}{} &  & \multicolumn{1}{c}{} &  & \tabularnewline
\end{tabular}
\par\end{centering}

\caption{\label{tab:SPP-active-interfaces-air}Analysis of SPP-activity of several metal ($\varepsilon_2$) surfaces irradiated in air ($\varepsilon_1 = 1$) at $\lambda=800$ nm and $400$ nm wavelengths. Meaning of the symbols: ${\color{ForestGreen}\protect\surd}:$ SPP are excited, ${\color{red}\times}$: SPP are not excited at the
interface. Bold font indicates interfaces where the prediction deviates from accepted theories. }
\end{table}

\begin{table}
\begin{centering}
\begin{tabular}{c|cc|cc}
\multicolumn{1}{c}{} & \multicolumn{2}{c}{} & \multicolumn{2}{c}{}\tabularnewline
\hline 
\hline 
\multirow{2}{*}{{\footnotesize{}Interface}} & \multicolumn{2}{c|}{{\footnotesize{}$\lambda=800$ nm}} & \multicolumn{2}{c}{{\footnotesize{}$\lambda=400$ nm}}\tabularnewline
 & {\footnotesize{}$\Im m\left(\varepsilon_{2}\right)=0$}  & {\footnotesize{}$\Im m\left(\varepsilon_{2}\right)\neq0$}  & {\footnotesize{}$\Im m\left(\varepsilon_{2}\right)=0$}  & {\footnotesize{}$\Im m\left(\varepsilon_{2}\right)\neq0$}\tabularnewline
\hline 
{\footnotesize{}SiO$_{2}$/Ag}  & \textbf{\textcolor{ForestGreen}{\footnotesize{}$\surd$}}\textcolor{ForestGreen}{{} } & \textbf{\textcolor{ForestGreen}{\footnotesize{}$\surd$}}\textcolor{ForestGreen}{{} } & \textbf{\textcolor{ForestGreen}{\footnotesize{}$\surd$}}\textcolor{ForestGreen}{{} } & \textbf{\textcolor{ForestGreen}{\footnotesize{}$\surd$}}\tabularnewline
{\footnotesize{}SiO$_{2}$/Al}  & \textbf{\textcolor{ForestGreen}{\footnotesize{}$\surd$}}\textcolor{ForestGreen}{{} } & \textbf{\textcolor{ForestGreen}{\footnotesize{}$\surd$}}\textcolor{ForestGreen}{{} } & \textbf{\textcolor{ForestGreen}{\footnotesize{}$\surd$}}\textcolor{ForestGreen}{{} } & \textbf{\textcolor{ForestGreen}{\footnotesize{}$\surd$}}\tabularnewline
\textbf{\textcolor{blue}{\footnotesize{}SiO$_{2}$/Au}}{\footnotesize{} } & \textbf{\textcolor{ForestGreen}{\footnotesize{}$\surd$}}\textcolor{ForestGreen}{{} } & \textbf{\textcolor{ForestGreen}{\footnotesize{}$\surd$}}\textcolor{ForestGreen}{{} } & \textbf{\textcolor{red}{\footnotesize{}$\times$}}{\footnotesize{} } & \textbf{\textcolor{ForestGreen}{\footnotesize{}$\surd$}}\tabularnewline
\textbf{\textcolor{blue}{\footnotesize{}SiO$_{2}$/Cr}}{\footnotesize{} } & \textbf{\textcolor{red}{\footnotesize{}$\times$}}{\footnotesize{} } & \textbf{\textcolor{ForestGreen}{\footnotesize{}$\surd$}}\textcolor{ForestGreen}{{} } & \textbf{\textcolor{ForestGreen}{\footnotesize{}$\surd$}}\textcolor{ForestGreen}{{} } & \textbf{\textcolor{ForestGreen}{\footnotesize{}$\surd$}}\tabularnewline
{\footnotesize{}SiO$_{2}$/Cu}  & \textbf{\textcolor{ForestGreen}{\footnotesize{}$\surd$}}\textcolor{ForestGreen}{{} } & \textbf{\textcolor{ForestGreen}{\footnotesize{}$\surd$}}\textcolor{ForestGreen}{{} } & \textbf{\textcolor{ForestGreen}{\footnotesize{}$\surd$}}\textcolor{ForestGreen}{{} } & \textbf{\textcolor{ForestGreen}{\footnotesize{}$\surd$}}\tabularnewline
{\footnotesize{}SiO$_{2}$/Fe } & \textbf{\textcolor{ForestGreen}{\footnotesize{}$\surd$}}\textcolor{ForestGreen}{{} } & \textbf{\textcolor{ForestGreen}{\footnotesize{}$\surd$}}\textcolor{ForestGreen}{{} } & \textbf{\textcolor{ForestGreen}{\footnotesize{}$\surd$}}\textcolor{ForestGreen}{{} } & \textbf{\textcolor{ForestGreen}{\footnotesize{}$\surd$}}\tabularnewline
\textbf{\textcolor{blue}{\footnotesize{}SiO$_{2}$/Nb}}{\footnotesize{} } & \textbf{\textcolor{ForestGreen}{\footnotesize{}$\surd$}}\textcolor{ForestGreen}{{} } & \textbf{\textcolor{ForestGreen}{\footnotesize{}$\surd$}}\textcolor{ForestGreen}{{} } & \textbf{\textcolor{red}{\footnotesize{}$\times$}}{\footnotesize{} } & \textbf{\textcolor{ForestGreen}{\footnotesize{}$\surd$}}\tabularnewline
{\footnotesize{}SiO$_{2}$/Ni}  & \textbf{\textcolor{ForestGreen}{\footnotesize{}$\surd$}}\textcolor{ForestGreen}{{} } & \textbf{\textcolor{ForestGreen}{\footnotesize{}$\surd$}}\textcolor{ForestGreen}{{} } & \textbf{\textcolor{ForestGreen}{\footnotesize{}$\surd$}}\textcolor{ForestGreen}{{} } & \textbf{\textcolor{ForestGreen}{\footnotesize{}$\surd$}}\tabularnewline
{\footnotesize{}SiO$_{2}$/Pt}  & \textbf{\textcolor{ForestGreen}{\footnotesize{}$\surd$}}\textcolor{ForestGreen}{{} } & \textbf{\textcolor{ForestGreen}{\footnotesize{}$\surd$}}\textcolor{ForestGreen}{{} } & \textbf{\textcolor{ForestGreen}{\footnotesize{}$\surd$}}\textcolor{ForestGreen}{{} } & \textbf{\textcolor{ForestGreen}{\footnotesize{}$\surd$}}\tabularnewline
{\footnotesize{}SiO$_{2}$/Ti}  & \textbf{\textcolor{ForestGreen}{\footnotesize{}$\surd$}}\textcolor{ForestGreen}{{} } & \textbf{\textcolor{ForestGreen}{\footnotesize{}$\surd$}}\textcolor{ForestGreen}{{} } & \textbf{\textcolor{ForestGreen}{\footnotesize{}$\surd$}}\textcolor{ForestGreen}{{} } & \textbf{\textcolor{ForestGreen}{\footnotesize{}$\surd$}}\tabularnewline
{\footnotesize{}SiO$_{2}$/V } & \textbf{\textcolor{red}{\footnotesize{}$\times$}}{\footnotesize{} } & \textbf{\textcolor{red}{\footnotesize{}$\times$}}{\footnotesize{} } & \textbf{\textcolor{ForestGreen}{\footnotesize{}$\surd$}}\textcolor{ForestGreen}{{} } & \textbf{\textcolor{ForestGreen}{\footnotesize{}$\surd$}}\tabularnewline
{\footnotesize{}SiO$_{2}$/W}  & \textbf{\textcolor{red}{\footnotesize{}$\times$}}{\footnotesize{} } & \textbf{\textcolor{red}{\footnotesize{}$\times$}}{\footnotesize{} } & \textbf{\textcolor{red}{\footnotesize{}$\times$}}{\footnotesize{} } & \textbf{\textcolor{red}{\footnotesize{}$\times$}}\tabularnewline
\hline 
\hline 
\multicolumn{1}{c}{} &  & \multicolumn{1}{c}{} &  & \tabularnewline
\end{tabular}
\par\end{centering}

\caption{\label{tab:SPP-active-interfaces-SiO2}Analysis of SPP-activity of several metal surfaces ($\varepsilon_2$) irradiated at $\lambda=800$ nm, and $400$ nm wavelengths through SiO$_2$ as covering medium [$\varepsilon_1($SiO$_2, \lambda = 800$ nm$) = 2.11$, $\varepsilon_1($SiO$_2, \lambda=400$ nm$) = 2.16$]. Meaning of the symbols: ${\color{ForestGreen}\protect\surd}:$ SPP are excited, ${\color{red}\times}$: SPP are not excited at the interface. Bold font indicates interfaces where the prediction deviates from accepted theories. }
\end{table}

In a second step, the same metals in contact with a semi-infinite dielectric medium (SiO$_2$) were addressed. Tab. \ref{tab:SPP-active-interfaces-SiO2} lists the corresponding results on SPP-activity. Again, with a few exceptions, agreement is found between the PMA and the generalized approach. The case of Nb is similar as in the analysis for air (compare Tabs. \ref{tab:SPP-active-interfaces-air} and \ref{tab:SPP-active-interfaces-SiO2}). Interestingly, at 400 nm wavelength, the PMA predicts no SPP-activity for the SiO$_2$/Au interface, while the more general treatment does. Also for SiO$_2$-covered Cr at 800 nm wavelength, the predictions of both theories do not match. 

In a third step, three most relevant metals (Ag, Au, Ti) in contact with a dielectric (air, Al$_2$O$_3$, SiO$_2$, TiO$_2$, ZnO) or semiconducting (GaAs, GaP, Ge, InP, Si, SiC) medium were analyzed regarding their SPP-activity at 800 nm and 400 nm wavelengths. Tab. \ref{tab:SPP-active-interfaces-Au} compiles the corresponding information and indicates that for Au and Ag at 800 nm, both models provide the same predictions on SPP-activity. However, for Ti at 800 nm and for all three metals at 400 nm, differences between predictions of both models can be observed. For Ti, the differences between those models arise for the significant contribution of the imaginary part of the dielectric permittivity [see Eq. (\ref{eq:ConditionsSPP1}), $\varepsilon_{2}(\mathrm{Ti}, \lambda=800$ nm$)= -2.85+i19.11$, $\varepsilon_{2}(\mathrm{Ti}, \lambda=400$ nm$)= -2.22+i6.66$]. 
For Au and Ag at 400 nm wavelength, the different model predictions arise from a similar origin (see Table \ref{tab:OpticalProperties}). 

\begin{table}
\begin{centering}
\begin{tabular}{c|cc|cc}
\multicolumn{1}{c}{} & \multicolumn{2}{c}{} & \multicolumn{2}{c}{}\tabularnewline
\hline 
\hline 
\multirow{2}{*}{{\footnotesize{}Interface}} & \multicolumn{2}{c|}{{\footnotesize{}$\lambda=800$ nm}} & \multicolumn{2}{c}{{\footnotesize{}$\lambda=400$ nm}}\tabularnewline
 & {\footnotesize{}$\Im m\left(\varepsilon_{2}\right)=0$}  & {\footnotesize{}$\Im m\left(\varepsilon_{2}\right)\neq0$}  & {\footnotesize{}$\Im m\left(\varepsilon_{2}\right)=0$}  & {\footnotesize{}$\Im m\left(\varepsilon_{2}\right)\neq0$} \tabularnewline
\hline 
{\footnotesize{}Air/Ag}  & \textcolor{ForestGreen}{\footnotesize{}$\surd$} & \textcolor{ForestGreen}{\footnotesize{}$\surd$} & \textcolor{ForestGreen}{\footnotesize{}$\surd$} & \textcolor{ForestGreen}{\footnotesize{}$\surd$}\tabularnewline
{\footnotesize{}Al$_{2}$O$_{3}$/Ag}  & \textcolor{ForestGreen}{\footnotesize{}$\surd$} & \textcolor{ForestGreen}{\footnotesize{}$\surd$} & \textcolor{ForestGreen}{\footnotesize{}$\surd$} & \textcolor{ForestGreen}{\footnotesize{}$\surd$}\tabularnewline
\textbf{\textcolor{blue}{\footnotesize{}GaAs/Ag}}{\footnotesize{} } & \textcolor{ForestGreen}{\footnotesize{}$\surd$} & \textcolor{ForestGreen}{\footnotesize{}$\surd$} & \textbf{\textcolor{red}{\footnotesize{}$\times$}} & \textcolor{ForestGreen}{\footnotesize{}$\surd$}\tabularnewline
\textbf{\textcolor{blue}{\footnotesize{}GaP/Ag}}{\footnotesize{} } & \textcolor{ForestGreen}{\footnotesize{}$\surd$} & \textcolor{ForestGreen}{\footnotesize{}$\surd$} & \textbf{\textcolor{red}{\footnotesize{}$\times$}} & \textcolor{ForestGreen}{\footnotesize{}$\surd$}\tabularnewline
\textbf{\textcolor{blue}{\footnotesize{}Ge/Ag}}{\footnotesize{} } & \textcolor{ForestGreen}{\footnotesize{}$\surd$} & \textcolor{ForestGreen}{\footnotesize{}$\surd$} & \textbf{\textcolor{red}{\footnotesize{}$\times$}} & \textcolor{ForestGreen}{\footnotesize{}$\surd$}\tabularnewline
\textbf{\textcolor{blue}{\footnotesize{}InP/Ag}}{\footnotesize{} } & \textcolor{ForestGreen}{\footnotesize{}$\surd$} & \textcolor{ForestGreen}{\footnotesize{}$\surd$} & \textbf{\textcolor{red}{\footnotesize{}$\times$}} & \textcolor{ForestGreen}{\footnotesize{}$\surd$}\tabularnewline
\textbf{\textcolor{blue}{\footnotesize{}Si/Ag}}{\footnotesize{} } & \textcolor{ForestGreen}{\footnotesize{}$\surd$} & \textcolor{ForestGreen}{\footnotesize{}$\surd$} & \textbf{\textcolor{red}{\footnotesize{}$\times$}} & \textcolor{ForestGreen}{\footnotesize{}$\surd$}\tabularnewline
\textbf{\textcolor{blue}{\footnotesize{}SiC/Ag}}{\footnotesize{} } & \textcolor{ForestGreen}{\footnotesize{}$\surd$} & \textcolor{ForestGreen}{\footnotesize{}$\surd$} & \textbf{\textcolor{red}{\footnotesize{}$\times$}} & \textcolor{ForestGreen}{\footnotesize{}$\surd$}\tabularnewline
{\footnotesize{}SiO$_{2}$/Ag}  & \textcolor{ForestGreen}{\footnotesize{}$\surd$} & \textcolor{ForestGreen}{\footnotesize{}$\surd$} & \textcolor{ForestGreen}{\footnotesize{}$\surd$} & \textcolor{ForestGreen}{\footnotesize{}$\surd$}\tabularnewline
\textbf{\textcolor{blue}{\footnotesize{}TiO$_{2}$/Ag}}{\footnotesize{} } & \textcolor{ForestGreen}{\footnotesize{}$\surd$} & \textcolor{ForestGreen}{\footnotesize{}$\surd$} & \textbf{\textcolor{red}{\footnotesize{}$\times$}} & \textcolor{ForestGreen}{\footnotesize{}$\surd$}\tabularnewline
\textbf{\textcolor{blue}{\footnotesize{}ZnO/Ag}}\textbf{\textcolor{blue}{{} }} & \textcolor{ForestGreen}{\footnotesize{}$\surd$} & \textcolor{ForestGreen}{\footnotesize{}$\surd$} & \textbf{\textcolor{red}{\footnotesize{}$\times$}} & \textcolor{ForestGreen}{\footnotesize{}$\surd$}\tabularnewline
\hline 
{\footnotesize{}Air/Au}  & \textcolor{ForestGreen}{\footnotesize{}$\surd$} & \textcolor{ForestGreen}{\footnotesize{}$\surd$} & \textcolor{ForestGreen}{\footnotesize{}$\surd$} & \textcolor{ForestGreen}{\footnotesize{}$\surd$}\tabularnewline
\textbf{\textcolor{blue}{\footnotesize{}Al$_{2}$O$_{3}$/Au}}{\footnotesize{} } & \textcolor{ForestGreen}{\footnotesize{}$\surd$} & \textcolor{ForestGreen}{\footnotesize{}$\surd$} & \textbf{\textcolor{red}{\footnotesize{}$\times$}} & \textcolor{ForestGreen}{\footnotesize{}$\surd$}\tabularnewline
{\footnotesize{}GaAs/Au } & \textcolor{ForestGreen}{\footnotesize{}$\surd$} & \textcolor{ForestGreen}{\footnotesize{}$\surd$} & \textbf{\textcolor{red}{\footnotesize{}$\times$}} & \textbf{\textcolor{red}{\footnotesize{}$\times$}}\tabularnewline
\textbf{\textcolor{blue}{\footnotesize{}GaP/Au}}{\footnotesize{} } & \textcolor{ForestGreen}{\footnotesize{}$\surd$} & \textcolor{ForestGreen}{\footnotesize{}$\surd$} & \textbf{\textcolor{red}{\footnotesize{}$\times$}} & \textcolor{ForestGreen}{\footnotesize{}$\surd$}\tabularnewline
{\footnotesize{}Ge/Au } & \textcolor{ForestGreen}{\footnotesize{}$\surd$} & \textcolor{ForestGreen}{\footnotesize{}$\surd$} & \textbf{\textcolor{red}{\footnotesize{}$\times$}} & \textbf{\textcolor{red}{\footnotesize{}$\times$}}\tabularnewline
{\footnotesize{}InP/Au}  & \textcolor{ForestGreen}{\footnotesize{}$\surd$} & \textcolor{ForestGreen}{\footnotesize{}$\surd$} & \textbf{\textcolor{red}{\footnotesize{}$\times$}} & \textbf{\textcolor{red}{\footnotesize{}$\times$}}\tabularnewline
\textbf{\textcolor{blue}{\footnotesize{}Si/Au}}{\footnotesize{} } & \textcolor{ForestGreen}{\footnotesize{}$\surd$} & \textcolor{ForestGreen}{\footnotesize{}$\surd$} & \textbf{\textcolor{red}{\footnotesize{}$\times$}} & \textcolor{ForestGreen}{\footnotesize{}$\surd$}\tabularnewline
\textbf{\textcolor{blue}{\footnotesize{}SiC/Au}}{\footnotesize{} } & \textcolor{ForestGreen}{\footnotesize{}$\surd$} & \textcolor{ForestGreen}{\footnotesize{}$\surd$} & \textbf{\textcolor{red}{\footnotesize{}$\times$}} & \textcolor{ForestGreen}{\footnotesize{}$\surd$}\tabularnewline
\textbf{\textcolor{blue}{\footnotesize{}SiO$_{2}$/Au}}{\footnotesize{} } & \textcolor{ForestGreen}{\footnotesize{}$\surd$} & \textcolor{ForestGreen}{\footnotesize{}$\surd$} & \textbf{\textcolor{red}{\footnotesize{}$\times$}} & \textcolor{ForestGreen}{\footnotesize{}$\surd$}\tabularnewline
\textbf{\textcolor{blue}{\footnotesize{}TiO$_{2}$/Au}}{\footnotesize{} } & \textcolor{ForestGreen}{\footnotesize{}$\surd$} & \textcolor{ForestGreen}{\footnotesize{}$\surd$} & \textbf{\textcolor{red}{\footnotesize{}$\times$}} & \textcolor{ForestGreen}{\footnotesize{}$\surd$}\tabularnewline
\textbf{\textcolor{blue}{\footnotesize{}ZnO/Au}}{\footnotesize{} } & \textcolor{ForestGreen}{\footnotesize{}$\surd$} & \textcolor{ForestGreen}{\footnotesize{}$\surd$} & \textbf{\textcolor{red}{\footnotesize{}$\times$}} & \textcolor{ForestGreen}{\footnotesize{}$\surd$}\tabularnewline
\hline 
{\footnotesize{}Air/Ti}  & \textcolor{ForestGreen}{\footnotesize{}$\surd$} & \textcolor{ForestGreen}{\footnotesize{}$\surd$} & \textcolor{ForestGreen}{\footnotesize{}$\surd$} & \textcolor{ForestGreen}{\footnotesize{}$\surd$}\tabularnewline
{\footnotesize{}Al$_{2}$O$_{3}$/Ti}  & \textcolor{ForestGreen}{\footnotesize{}$\surd$} & \textcolor{ForestGreen}{\footnotesize{}$\surd$} & \textcolor{ForestGreen}{\footnotesize{}$\surd$} & \textcolor{ForestGreen}{\footnotesize{}$\surd$}\tabularnewline
\textbf{\textcolor{blue}{\footnotesize{}GaAs/Ti}}{\footnotesize{} } & \textcolor{red}{\footnotesize{}$\times$} & \textcolor{ForestGreen}{\footnotesize{}$\surd$} & \textcolor{red}{\footnotesize{}$\times$} & \textcolor{red}{\footnotesize{}$\times$}\tabularnewline
\textbf{\textcolor{blue}{\footnotesize{}GaP/Ti}}{\footnotesize{} } & \textcolor{red}{\footnotesize{}$\times$}{\footnotesize{} } & \textcolor{ForestGreen}{\footnotesize{}$\surd$} & \textcolor{red}{\footnotesize{}$\times$} & \textcolor{ForestGreen}{\footnotesize{}$\surd$}\tabularnewline
\textbf{\textcolor{blue}{\footnotesize{}Ge/Ti}}{\footnotesize{} } & \textcolor{red}{\footnotesize{}$\times$}{\footnotesize{} } & \textcolor{ForestGreen}{\footnotesize{}$\surd$} & \textcolor{red}{\footnotesize{}$\times$} & \textcolor{red}{\footnotesize{}$\times$}\tabularnewline
\textbf{\textcolor{blue}{\footnotesize{}InP/Ti}}{\footnotesize{} } & \textcolor{red}{\footnotesize{}$\times$}{\footnotesize{} } & \textcolor{ForestGreen}{\footnotesize{}$\surd$} & \textcolor{red}{\footnotesize{}$\times$} & \textcolor{red}{\footnotesize{}$\times$}\tabularnewline
\textbf{\textcolor{blue}{\footnotesize{}Si/Ti}}{\footnotesize{} } & \textcolor{red}{\footnotesize{}$\times$}{\footnotesize{} } & \textcolor{ForestGreen}{\footnotesize{}$\surd$} & \textcolor{red}{\footnotesize{}$\times$} & \textcolor{ForestGreen}{\footnotesize{}$\surd$}\tabularnewline
\textbf{\textcolor{blue}{\footnotesize{}SiC/Ti}}{\footnotesize{} } & \textbf{\textcolor{red}{\footnotesize{}$\times$ }}{\footnotesize{} } & \textcolor{ForestGreen}{\footnotesize{}$\surd$} & \textbf{\textcolor{red}{\footnotesize{}$\times$}} & \textcolor{ForestGreen}{\footnotesize{}$\surd$}\tabularnewline
{\footnotesize{}SiO$_{2}$/Ti}  & \textcolor{ForestGreen}{\footnotesize{}$\surd$}\textcolor{ForestGreen}{{} } & \textcolor{ForestGreen}{\footnotesize{}$\surd$} & \textcolor{ForestGreen}{\footnotesize{}$\surd$} & \textcolor{ForestGreen}{\footnotesize{}$\surd$}\tabularnewline
\textbf{\textcolor{blue}{\footnotesize{}TiO$_{2}$/Ti}}\textbf{\textcolor{blue}{{} }} & \textcolor{red}{\footnotesize{}$\times$}{\footnotesize{} } & \textcolor{ForestGreen}{\footnotesize{}$\surd$} & \textcolor{red}{\footnotesize{}$\times$} & \textcolor{ForestGreen}{\footnotesize{}$\surd$}\tabularnewline
\textbf{\textcolor{blue}{\footnotesize{}ZnO/Ti}}{\footnotesize{} } & \textcolor{ForestGreen}{\footnotesize{}$\surd$} & \textcolor{ForestGreen}{\footnotesize{}$\surd$} & \textcolor{red}{\footnotesize{}$\times$} & \textcolor{ForestGreen}{\footnotesize{}$\surd$}\tabularnewline
\hline 
\hline 
\multicolumn{1}{c}{} &  & \multicolumn{1}{c}{} &  & \tabularnewline
\end{tabular}
\par\end{centering}

\caption{\label{tab:SPP-active-interfaces-Au}Analysis of SPP-activity of three metal surfaces (Ag, Au, Ti) irradiated through 
different dielectric or semiconducting media at $\lambda=800$ nm and
$400$ nm wavelengths. Meaning of the symbols: ${\color{ForestGreen}\protect\surd}:$
SPP are excited, ${\color{red}\times}$: SPP are not excited at the
interface. Bold font indicates interfaces where the prediction deviates
from accepted theories. }
\end{table}

\section{Period of Surface Plasmon Polaritons in lossy materials \label{sec:Period}}

\subsection{Modeling the plasmon period}

The spatial period $\Lambda$ of the electromagnetic field can be calculated from the complex-valued SPP wavenumber $\beta$ [using Eq. (\ref{eq:BetaDefinition})] via \cite{Bell1973,Ionin2014}

\begin{equation}
\Lambda=\frac{2\pi}{\Re e\left(\beta\right)}\label{eq:SPPperiod}, 
\end{equation}

with 
\[
\Re e\left(\beta\right)=\frac{\omega}{2c}\,\sqrt{2\,\left[F_{1}^{2}+F_{2}^{2}\right]^{1/2}+2F_{1}},
\]
where 
\[
F_{1}=\left(\varepsilon_{1}'\varepsilon_{2}'\underbrace{-\varepsilon_{1}''\varepsilon_{2}''}\right)\left(\varepsilon_{1}'+\varepsilon_{2}'\right)F_{0}+\underbrace{\left(\varepsilon_{1}''\varepsilon_{2}'+\varepsilon_{1}'\varepsilon_{2}''\right)\left(\varepsilon_{1}''+\varepsilon_{2}''\right)F_{0}},
\]
 
\[
F_{2}=\underbrace{\left(\varepsilon_{1}''\varepsilon_{2}'+\varepsilon_{1}'\varepsilon_{2}''\right)\left(\varepsilon_{1}'+\varepsilon_{2}'\right)F_{0}-\left(\varepsilon_{1}'\varepsilon_{2}'-\varepsilon_{1}''\varepsilon_{2}''\right)\left(\varepsilon_{1}''+\varepsilon_{2}''\right)F_{0}},
\]
 and $F_{0}=\left[\left(\varepsilon_{1}'+\varepsilon_{2}'\right)^{2}\underbrace{+\left(\varepsilon_{1}''+\varepsilon_{2}''\right)^{2}}\right]^{-1}$. 
 The terms underlined with a curly brackets can be neglected if the PMA is used, i.e., if $\varepsilon''_{1}=0$ and $\varepsilon''_{2}=0$. 
From Eq. (\ref{eq:SPPperiod}), it is obvious that the imaginary parts of both media 
can significantly affect the SPP periods. 

For the PMA ($\varepsilon''_{1}=0$ and $\varepsilon''_{2}=0$), Eq. (\ref{eq:SPPperiod}) reduces to the expression \cite{Bonch-Bruevich1992}
\begin{equation}
\Lambda_{\mathrm{PMA}} = \frac{2 \pi}{\Re e (\beta)} = \frac{\lambda}{\sqrt{\frac{\varepsilon'_1 \varepsilon'_2}{\varepsilon'_1 + \varepsilon'_2}}}\label{eq:PMAperiod}.
\end{equation}
However, it should be underlined that Eq. (\ref{eq:PMAperiod}) should not be applied to lossy materials. 

\subsection{Exploration of plasmon-active material combinations}

In this section, different SPP-active combinations of materials are systematically explored in terms of their SPP period using Eq. (\ref{eq:SPPperiod}). 

\begin{figure}
\begin{centering}
\includegraphics[width=8cm]{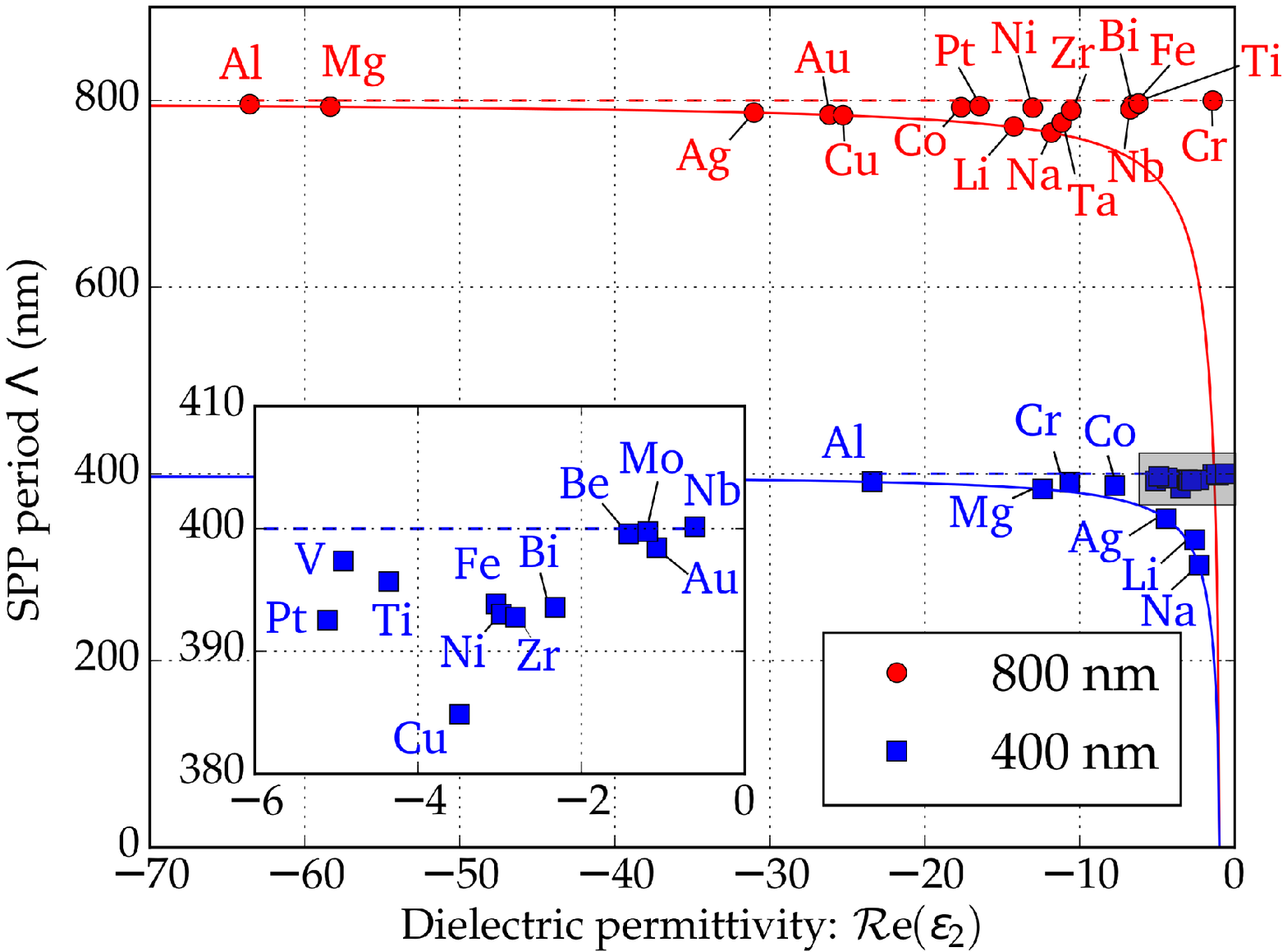}
\par\end{centering}

\caption{\label{fig:Resonant-SPP-period-Air} SPP period for several metals irradiated in air at 800 nm (red circles) and 400 nm (blue squares) wavelengths \revadd{calculated with Eq. (\ref{eq:SPPperiod}). }The entities $\Lambda = \lambda$ are shown as horizontal dashed lines for both wavelengths. \revadd{The solid lines represent the periods calculated using the PMA [Eq. (\ref{eq:PMAperiod})].} The inset displays a magnification of the area marked by the grey rectangle. }
\end{figure}

Fig. \ref{fig:Resonant-SPP-period-Air} shows for metals being SPP-active in air the spatial period $\Lambda$ as a function of their real parts of $\varepsilon_2$. 
\revadd{At 800 nm wavelength, the SPP-periods calculated by Eq. (\ref{eq:SPPperiod}) (discrete data points) match within $5\%$ to the irradiation wavelength $\lambda$ - as indicated by a \revadd{horizontal red dashed line}. The solution based on the PMA (Eq. \ref{eq:PMAperiod}) with $\varepsilon_1 = 1$ is shown as red solid line. By comparison, it is obvious that the PMA (Eq. \ref{eq:PMAperiod}) should not be applied to lossy materials. 
At 400 nm wavelength, the SPP-periods more strongly depend on the material. A reduction of $\Lambda$ by up to $25 \%$, when compared to the wavelength $\lambda$, is observed for Ag, Li, and Na - see the blue horizontal dashed line, with a similar mismatch to the PMA curve (blue solid line).}

\begin{figure}
\begin{centering}
\includegraphics[width=8cm]{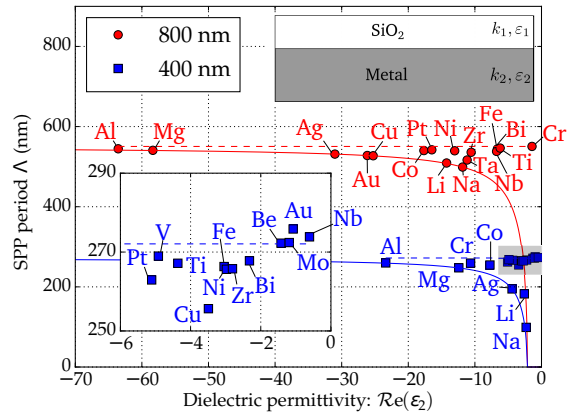}
\par\end{centering}
\caption{\label{fig:Resonant-SPP-period-SiO2} SPP period for several metals irradiated through SiO$_2$ ($\varepsilon_1$) at 800 nm (red circles) and 400 nm (blue squares) wavelengths \revadd{calculated with Eq. (\ref{eq:SPPperiod}).} The entities $\Lambda = \lambda / \Re e\sqrt{\varepsilon_1}$ are shown as horizontal dashed lines for both wavelengths. \revadd{The solid lines represent the periods calculated using the PMA [Eq. (\ref{eq:PMAperiod})] with a fixed $\varepsilon_1$.} The inset displays a magnification of the area marked by the grey rectangle.}
\end{figure}

Fig. \ref{fig:Resonant-SPP-period-SiO2} shows the results of analogous analyses for SPP-active metals ($\varepsilon_2$) in contact with SiO$_2$ ($\varepsilon_1$). At 800 nm wavelength, the SPP-periods match within $8\%$ to the wavelength in SiO$_2$ (i.e., above the metal surface, $\lambda / \Re e \sqrt{\varepsilon_1}$) \cite{Siegman1986} - as indicated by a \revadd{horizontal red dashed line}. At 400 nm wavelength, for most metals, the deviations from the latter expression also stay below $10 \%$ \revadd{(compared to the horizontal  blue dashed line). }In contrast, for Ag, Li and Na, larger differences up to $65 \%$ are evident. \revadd{For both wavelengths, the data points and the periods calculated using the PMA [Eq. (\ref{eq:PMAperiod}), solid lines] match only for specific materials. }

\begin{figure}
\begin{centering}
\includegraphics[width=8cm]{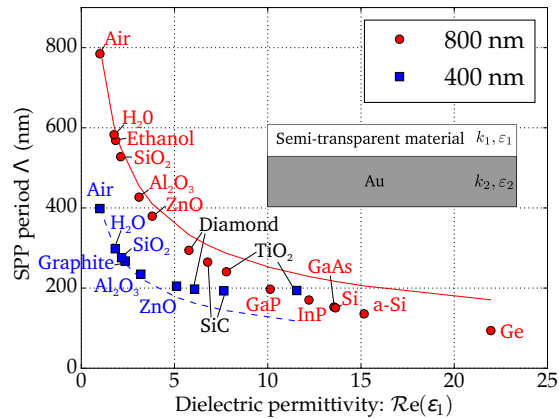}
\par\end{centering}

\caption{\label{fig:Resonant-SPP-period-Au} SPP period for Au irradiated through various types of semi-transparents materials at 800 nm (red circles) or 400 nm (blue squares) wavelengths. The entities $\Lambda = \lambda / \Re e \sqrt{\varepsilon_1}$ are shown as lines for both wavelengths.}
\end{figure}

In order to understand the role of the overlayer, the SPP-period is investigated at the interface between Au ($\varepsilon_2$) and various semi-transparent materials ($\varepsilon_1$). The results are provided in Fig. \ref{fig:Resonant-SPP-period-Au}. They indicate that, again, the SPP-periods strongly depends on $\varepsilon_1$. Specifically, a reasonably good agreement is found between the SPP-period $\Lambda$ and the wavelength in the covering medium $\lambda / \Re e \sqrt{\varepsilon_1}$. 
This can be seen by the red solid ($\lambda = 800$ nm) and blue dashed ($\lambda = 400$ nm) lines, which quantitatively agree with the data points for small values of $\Re e (\varepsilon_1)$. The remarkable deviations at larger values of $\Re e (\varepsilon_1)$ arise from the PMA. 

\begin{figure}
\begin{centering}
\includegraphics[width=8cm]{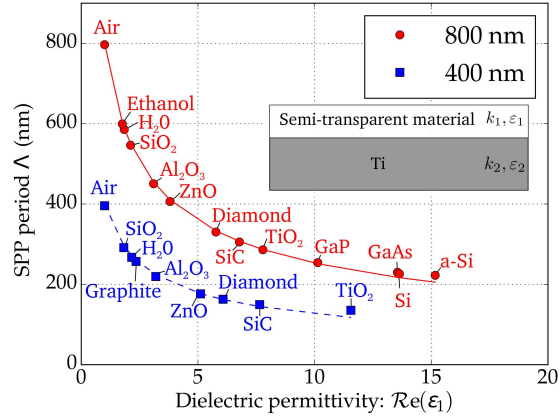}
\par\end{centering}

\caption{\label{fig:Resonant-SPP-period-Ti} SPP period for Ti irradiated through various types of semi-transparents materials at 800 nm (red circles) or 400 nm (blue squares) wavelengths. The entities $\Lambda = \lambda / \Re e \sqrt{\varepsilon_1}$ are shown as lines for both wavelengths. }
\end{figure}

Fig. \ref{fig:Resonant-SPP-period-Ti} shows analogous results for the case of a Ti substrate ($\varepsilon_2$) covered by the same set of materials ($\varepsilon_1$). Again, the lines represent the wavelengths in the covering medium $\lambda / \Re e \sqrt{\varepsilon_1}$, which are in reasonable agreement with the data points. 

It should be underlined that the conclusions drawn in this work 
strongly depend on the quality of the optical data used. All parameters 
affecting the dielectric permittivities may influence the predictions, i.e., 
the temperature \cite{Landolt}, high-intensity illumination \cite{Bonse2009,Derrien2013,Bevillon2015a},
strong external fields, etc. Moreover, we have assumed that the SPP-active interface is surrounded 
by two semi-infinite half-spaces. This assumption may break down when additional 
interfaces fall within the vertical extend of SPP field. Then another model has 
to be used which considers the coherent coupling of SPP at several interfaces \cite{Ward1975a,Derrien2014a}.

\section{Lifetime of Surface Plasmon Polaritons in lossy materials}

So far, we have described the Surface Plasmon Polaritons within the frame of steady-state conditions. Once the driving external radiation field is turned off, the SPPs may exist for a certain duration called lifetime $\tau_{\mbox{\small SPP}}$ in the following. Upon internal damping, via electron collisions, re-radiation of light to the far field \cite{Mueller2010} via scattering on defects, etc., SPP typically vanish on a sub-picosecond timescale. However, the exact lifetime strongly depends on the material and irradiation parameters. 

In this section, two different models of SPP lifetime are introduced and applied to 
the materials Ag and Ti in a wide range of irradiation wavelengths (from UV to IR). 
\revadd{Two different models from literature will be compared \cite{Raether1986, Hohenau2008}. }

\revadd{The SPP propagation length $L_{SPP}$ ($1/e$-decay of the intensity) is defined by \cite{Hohenau2008}
\begin{equation}
L_{SPP} := [2\times\Im m{(\beta)}]^{-1}.\label{eq:PropagationLength} \label{eq:Lspp1}
\end{equation}
Using the group velocity ($v_{g}$) and the lifetime ($\tau_{SPP}$) of SPPs, their propagation length can be approximated by: 
\begin{equation}
L_{SPP} \sim v_g \times \tau_{SPP} \label{eq:Lspp2}
\end{equation}
The group velocity is given by the definition
\begin{equation}
v_{g} :=\frac{d \omega}{d [\Re e (\beta)]} = \frac{c}{n_{SP} (\lambda) - \lambda \frac{d n_{SP}}{d\lambda} }\label{eq:GroupVelocity}
\end{equation}
with $n_{SP} (\lambda) = \Re e \sqrt{\frac{\varepsilon_1 (\lambda) \varepsilon_2 (\lambda)}{\varepsilon_1 (\lambda) + \varepsilon_2 (\lambda)}}$  \cite{Hohenau2008, Jackson1999}.}

\revadd{Combining Eqs. (\ref{eq:Lspp1})-(\ref{eq:GroupVelocity}) results in an SPP lifetime of
\begin{equation}
\tau_{SPP} \sim \frac{1}{2 \times \Im m (\beta) \times v_g}. \label{eq:LifeTimeHohenau}
\end{equation}}

\revadd{Another expression of the SPP lifetime was provided by Raether \cite{Raether1986} to be: 
\begin{equation}
\tau_{SPP}=\frac{1}{ 2 \times \Im m\left( \omega_{SPP}\right)}\label{eq:LifeTimeRaether},
\end{equation}}
where
\begin{equation}
\Im m\left(\omega_{SPP}\right)= \frac{c}{2} \times \frac{\Re e\left(\beta\right) \Im m\left(\epsilon_{2}\right)}{ \Re e\left(\varepsilon_{2}\right)^{2}}\times\frac{\Re e\left(\varepsilon_{1}\right) \Re e (\varepsilon_{2} ) }{\Re e\left(\varepsilon_{1}\right)+ \Re e (\varepsilon_{2})}.\label{eq:Damping}
\end{equation}

In the following two subsections, we apply \revadd{and compare the two Eqs. (\ref{eq:LifeTimeHohenau}) and (\ref{eq:LifeTimeRaether})} to the well-known case of air-Ag interface (Example 1, \revadd{a material properly described by the PMA)} and the less studied air-Ti interface \revadd{(Example 2, representing a lossy material). }

\subsection*{Example 1 \revadd{("lossless" material)}: air - Ag interface}

\begin{figure}
\begin{centering}
\includegraphics[width=8cm]{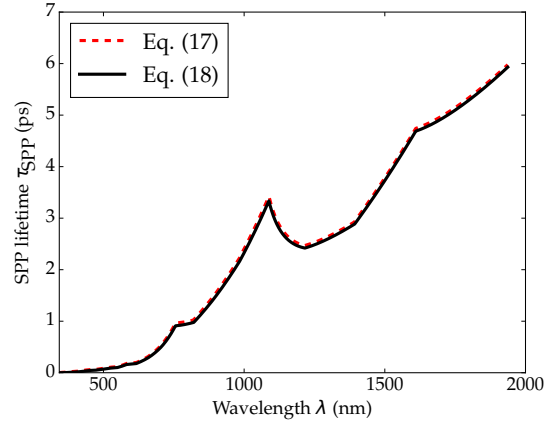}
\par\end{centering}

\caption{\label{fig:SPP-lifetime-Ag} SPP lifetime $\tau_{\mbox{\small SPP}}$ at the air - Ag interface as a function of the irradiation wavelength $\lambda$. \revadd{Eqs. (\ref{eq:LifeTimeHohenau}) and (\ref{eq:LifeTimeRaether}) were used to compute the red dashed and black solid lines}, respectively.} 
\end{figure}

Figure \ref{fig:SPP-lifetime-Ag} presents the SPP lifetime
$\tau_{\mbox{\small SPP}}$ as a function of \revadd{the irradiation
wavelength} $\lambda$ \revadd{(350 nm} - 2 $\mu$m). The two curves shown in Fig. \ref{fig:SPP-lifetime-Ag} are based
on \revadd{Eqs. (\ref{eq:LifeTimeHohenau}) and
(\ref{eq:LifeTimeRaether}) }along with the optical data from Ref.
\cite{Johnson1972}. Both curves exhibit an excellent agreement
featuring lifetimes up to \revadd{6} ps. At 400 nm wavelength,
$\tau_{\mbox{\small SPP}}$ accounts to \revadd{14 fs [Eq.
(\ref{eq:LifeTimeRaether})] and 27 fs [Eq.
(\ref{eq:LifeTimeHohenau})],} while at 800 nm wavelength,
both equations provide a very similar value of \revadd{1 ps.} At 400 nm
wavelength, $\tau_{\mbox{\small SPP}}$ agrees \revadd{reasonably well} with the
experiments of \emph{Kubo et al.} \cite{Kubo2005} who observed
the SPP decay within 50 fs for the air-Ag interface. It should
be underlined that the SPP lifetime strongly depends on the
irradiation wavelength, varying by more than two orders of
magnitude here.

\subsection*{Example 2 \revadd{(lossy material)}: air - Ti interface}

\begin{figure}
\begin{centering}
\includegraphics[width=8cm]{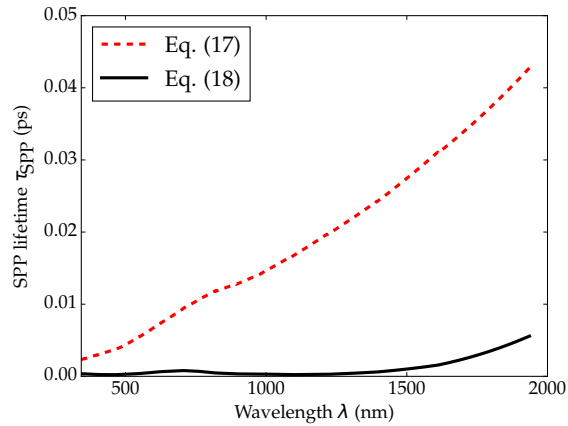}
\par\end{centering}

\caption{\label{fig:SPP-lifetime-Ti}SPP lifetime $\tau_{\mbox{\small SPP}}$ at the air - Ti interface as a function of the irradiation wavelength $\lambda$. \revadd{Eqs. (\ref{eq:LifeTimeHohenau}) and (\ref{eq:LifeTimeRaether}) were used to compute the red dashed and black solid lines}, respectively.}
\end{figure}

Figure \ref{fig:SPP-lifetime-Ti} displays 
the SPP lifetime $\tau_{\mbox{\small SPP}}$ at the air -
Ti interface, as a function of \revadd{the irradiation
wavelength $\lambda$.}
\revadd{Again, the} two curves shown in Fig.
\ref{fig:SPP-lifetime-Ti} are calculated from Eqs.
\revadd{(\ref{eq:LifeTimeHohenau}) and (\ref{eq:LifeTimeRaether})} using
the optical data from Ref. \cite{Johnson1972}. In contrast to
Example 1 (Ag), the two curves are remarkably different for the
air-Ti interface. The model based on Eq. (\ref{eq:LifeTimeRaether})
provides values which are smaller by more than one order of
magnitude when compared to the ones provided by Eq.
(\ref{eq:LifeTimeHohenau}). Specifically, at 400 nm
wavelength, $\tau_{\mbox{\small SPP}}$ \revadd{accounts to $0.3$ fs [Eq.
(\ref{eq:LifeTimeRaether})] versus $3$ fs [Eq.
(\ref{eq:LifeTimeHohenau})].} At 800 nm wavelength, it
accounts to \revadd{$0.5$ fs [Eq. (\ref{eq:LifeTimeRaether})] versus $11$ fs [Eq.
(\ref{eq:LifeTimeHohenau})]. The values calculated by Eq.
(\ref{eq:LifeTimeRaether})} appear unreasonably small here, as they are
smaller than the expected Drude relaxation time of the electrons
\cite{Ordal1985}. It is important to emphasize that the two
models result in very different predictions for Ti.

\section{Conclusion}

In this work, the properties of Surface Plasmon Polaritons at the
interface between two lossy materials have been studied. 
Our mathematical analysis allowed to identify terms which should not be 
neglected in the mathematical description of SPPs on lossy materials. 
This rigorous approach was applied to numerous material combinations (dielectric/metal, semiconductor/metal), providing a generalized criterion for SPP excitation along with the quantification of the SPP periods. 
For Ag and Ti surfaces interfaced with air, the lifetimes of the SPP were reported in a wide spectral range between \revadd{350 nm} and 2 $\mu$m wavelength pointing out again that plasmonic loss mechanisms should be carefully taken into consideration. 

\section*{Acknowledgments}

Fruitful discussions and support from Prof.~N.M.~Bulgakova are gratefully acknowledged. Critical discussions of the mathematical calculations with \revadd{Dr.~Y. Levy} are also acknowledged. \revadd{The authors also acknowledge the fruitful remarks of one of the anonymous referees.}

This research was supported by the state budget of the Czech Republic (project HiLASE: Superlasers for the real world: LO1602). 
T.J.-Y.D. acknowledges the support of Marie Sk\l{}odowska Curie Actions
(MSCA) Individual Fellowship of the European's Union's (EU) Horizon 2020 Programme under grant agreement ''QuantumLaP'' No. 657424.
 
\section*{References}

%\bibliographystyle{/media/thibault/PHD/Travail/Berlin/Publications/DoublePulseSi/osajnl}
%\bibliography{/media/thibault/PHD/Travail/Bibliography/bibliographie_lue}

\begin{thebibliography}{10}
\newcommand{\enquote}[1]{``#1''}

\bibitem{Maier2007}
S.~A. Maier, Plasmonics, Fundamentals and Applications, Springer, 2007.

\bibitem{MacDonald2009}
K.~F. MacDonald, Z.~L. Samson, M.~I. Stockman, N.~I. Zheludev,
  \href{http://www.nature.com/doifinder/10.1038/nphoton.2008.249}{Ultrafast
  active plasmonics}, Nat. Photonics 3~(1) (2009) 55--58.

\bibitem{Atwater2010}
H.~A. Atwater, A.~Polman, Plasmonics for improved photovoltaic devices, Nat.
  Mater. 9 (2010) 205--213.

\bibitem{Otto1974}
A.~Otto,
  \href{http://link.springer.com/chapter/10.1007/BFb0108460}{Experimental
  investigation of surface polaritons on plane interfaces}, in:
  Festk\"{o}rperprobleme 14, Springer, 1974, pp. 1--37.

\bibitem{Bell1975a}
R.~J. Bell, R.~W.~A. Jr., C.~Ward, I.~L.Tyler, Introductory theory for surface
  electromagnetic wave spectroscopy, Surf. Sci. 48 (1975) 253--287.

\bibitem{Kabashin2009}
A.~Kabashin, P.~Evans, S.~Pastkovsky, W.~Hendren, G.~Wurtz, R.~Atkinson,
  R.~Pollard, V.~Podolskiy, A.~Zayats, Plasmonic nanorod metamaterials for
  biosensing, Nat. Mater. 8 (2009) 867--871.

\bibitem{Berini2011}
P.~Berini, I.~De~Leon,
  \href{http://www.nature.com/doifinder/10.1038/nphoton.2011.285}{Surface
  plasmon polariton amplifiers and lasers}, Nat. Photonics 6~(1) (2011) 16--24.

\bibitem{Park2011}
I.-Y. Park, S.~Kim, J.~Choi, D.-H. Lee, Y.-J. Kim, M.~F. Kling, M.~I. Stockman,
  S.-W. Kim,
  \href{http://www.nature.com/doifinder/10.1038/nphoton.2011.258}{Plasmonic
  generation of ultrashort extreme-ultraviolet light pulses}, Nat. Photonics
  5~(11) (2011) 677--681.

\bibitem{Barnes2003}
W.~Barnes, A.~Dereux, T.~Ebbesen, Surface plasmon subwavelength optics, Nature
  424 (2003) 824--830.

\bibitem{Liu2008a}
H.~Liu, P.~Lalanne,
  \href{http://www.nature.com/doifinder/10.1038/nature06762}{Microscopic theory
  of the extraordinary optical transmission}, Nature 452~(7188) (2008)
  728--731.

\bibitem{Shalaev2007}
V.~M. Shalaev, Optical negative-index metamaterials, Nat. Photonics 1 (2007)
  41--48.

\bibitem{Kim2008b}
S.~Kim, J.~Jin, Y.-J. Kim, I.-Y. Park, Y.~Kim, S.-W. Kim,
  \href{http://www.nature.com/doifinder/10.1038/nature07012}{High-harmonic
  generation by resonant plasmon field enhancement}, Nature 453~(7196) (2008)
  757--760.

\bibitem{Genevet2010}
P.~Genevet, J.-P. Tetienne, E.~Gatzogiannis, R.~Blanchard, M.~A. Kats, M.~O.
  Scully, F.~Capasso,
  \href{http://pubs.acs.org/doi/abs/10.1021/nl102747v}{Large {Enhancement} of
  {Nonlinear} {Optical} {Phenomena} by {Plasmonic} {Nanocavity} {Gratings}},
  Nano Lett. 10~(12) (2010) 4880--4883.

\bibitem{Purvis2013}
M.~A. Purvis, V.~N. Shlyaptsev, R.~Hollinger, C.~Bargsten, A.~Pukhov,
  A.~Prieto, Y.~Wang, B.~M. Luther, L.~Yin, S.~Wang, J.~J. Rocca,
  \href{http://www.nature.com/doifinder/10.1038/nphoton.2013.217}{Relativistic
  plasma nanophotonics for ultrahigh energy density physics}, Nat. Photonics
  7~(10) (2013) 796--800.

\bibitem{Pettit1975}
R.~B. Pettit, J.~Silcox, R.~Vincent,
  \href{http://journals.aps.org/prb/abstract/10.1103/PhysRevB.11.3116}{Measurement
  of surface-plasmon dispersion in oxidized aluminum films}, Phys. Rev. B
  11~(8) (1975) 3116.

\bibitem{Raether1986}
H.~Raether, {S}urface {P}lasmons on {S}mooth and {R}ough {S}urfaces and on
  {G}ratings, Springer-Verlag, 1986.

\bibitem{Bell1973}
R.~J. Bell, R.~W. Alexander, W.~F. Parks, G.~Kovener, Surface excitations in
  absorbing media, Opt. Commun. 8~(2) (1973) 147.

\revadd{\bibitem{Alexander1974a}
R.~W. Alexander, G.~S. Kovener, and R.~J. Bell, Dispersion curves for
  surface electromagnetic waves with damping, Phys. Rev. Lett. 32,
  154 (1974).}

\bibitem{Ward1974}
C.~A. Ward, R.~J. Bell, R.~W. Alexander, G.~S. Kovener, I.~Tyler,
  \href{http://www.osapublishing.org/abstract.cfm?uri=ao-13-10-2378}{Surface
  electromagnetic waves on metals and polar insulators: some comments}, Appl.
  Opt. 13~(10) (1974) 2378--2381.

\bibitem{Ward1975a}
C.~A. Ward, R.~W. Alexander, R.~J. Bell,
  \href{http://journals.aps.org/prb/abstract/10.1103/PhysRevB.12.3293}{Surface
  electromagnetic waves on layered systems with damping}, Phys. Rev. B 12~(8)
  (1975) 3293.

\bibitem{Kovener1976}
G.~S. Kovener, R.~W. Alexander~Jr, R.~J. Bell,
  \href{http://journals.aps.org/prb/abstract/10.1103/PhysRevB.14.1458}{Surface
  electromagnetic waves with damping. {I}. {Isotropic} media}, Phys. Rev. B
  14~(4) (1976) 1458.

\bibitem{Kovener1977}
G.~S. Kovener, R.~W. Alexander~Jr, I.~L. Tyler, R.~J. Bell,
  \href{http://journals.aps.org/prb/abstract/10.1103/PhysRevB.15.5877}{Surface
  electromagnetic waves with damping. {II}. {Anisotropic} media}, Phys. Rev. B
  15~(12) (1977) 5877.

\revadd{\bibitem{Ritchie1977}
R.~H. Ritchie, R.~N. Hamm, M.~W. Williams, and E.~T. Arakawa,
  Dispersion relations of elementary excitations when damping is
  present, Phys. Status Solidi 84 (1977) 367.}

\bibitem{Laglois1978}
J.~Laglois, B.~Fischer, Dispersion theory of surface-exciton polaritons, Phys.
  Rev. B 17 (1978) 3814.

\bibitem{Reinisch1979}
R.~Reinisch, Theory of nonlinear excitation of surface polaritons in lossy
  media, Surf. Sci. 85 (1979) 432--456.

\bibitem{Boardman1984}
A.~D. Boardman, P.~Egan,
  \href{http://www.edpsciences.org/10.1051/jphyscol:1984526}{The influence of
  collisional damping on surface plasmon-polariton dispersion}, J. Phys.
  Colloques 45~(C5) (1984) 179--190.

\revadd{\bibitem{Norrman2013}
A.~Norrman, T.~Set\"{a}l\"{a}, and A.~T. Fridberg, Exact
  surface-plasmon polariton solutions at a lossy interface, Opt. Lett.
  38 (2013) 7.}

\revadd{\bibitem{Lee2013a}
H.-I. Lee and J.~Mok, Electromagnetic-energy conservation for media
  with metallic constituents with special attention to damped waves, J. Opt.
  15 (2013) 035002.}

\revadd{\bibitem{Martinez-Herrero2015}
R.~Martinez-Herrero, A.~Garcia-Ruiz, and A.~Manjavacas, Parametric
  characterization of surface plasmon polaritons at a lossy interface, Opt.
  Express 23 (2015) 28574--28583.}

\bibitem{Zayats2005}
A.~Zayats, I.~Smolyaninoy, A.~Maradudin, Nano-optics of surface plasmon
  polaritons, Phys. Rep. 408 (2005) 131--314.

\bibitem{Bonch-Bruevich1992}
A.~M. Bonch-Bruevich, M.~N. Libenson, V.~S. Makin, V.~A. Trubaev, Surface
  electromagnetic waves in optics, Opt. Eng. 31~(4) (1992) 718--730.

\bibitem{Palik1985}
E.~D. Palik, Handbook of Optical Constants of Solids, Academic Press, 1985.

\bibitem{Ionin2014}
A.~Ionin, S.~Kudryashov, S.~Makarov, A.~Rudenko, P.~Saltuganov, L.~Seleznev,
  D.~Sinitsyn, E.~Sunchugasheva,
  \href{http://linkinghub.elsevier.com/retrieve/pii/S0169433213023027}{Femtosecond
  laser fabrication of sub-diffraction nanoripples on wet al surface in
  multi-filamentation regime: High optical harmonics effects?}, Appl. Surf.
  Sci. 292 (2014) 678--681.

\bibitem{Siegman1986}
A.~E. Siegman, P.~M. Fauchet, Stimulated {W}ood's anomalies on
  laser-illuminated surfaces, IEEE J. Quantum Electron. QE-22 (1986) 1384.

\bibitem{Landolt}
H.~H. Landolt, R.~B\"{o}rnstein, Landolt - B\"{o}rnstein Numerical Data and
  Functional Relationships, Springer-Verlag.

\bibitem{Bonse2009}
J.~Bonse, A.~Rosenfeld, J.~Kr\"uger, On the role of surface plasmon polaritons
  in the formation of laser-induced periodic surface structures upon
  irradiation of silicon by femtosecond-laser pulses, J. Appl. Phys. 106 (2009)
  104910.

\bibitem{Derrien2013}
T.~J.-Y. Derrien, T.~E. Itina, R.~Torres, T.~Sarnet, M.~Sentis, Possible
  surface plasmon polariton excitation under femtosecond laser irradiation of
  silicon, J. Appl. Phys. 114 (2013) 083104.

\bibitem{Bevillon2015a}
E.~B\'{e}villon, J.~P. Colombier, V.~Recoules, H.~Zhang, C.~Li, R.~Stoian,
  Ultrafast {Surface} {Plasmonic}
  {Switch} in {Non}-{Plasmonic} {Metals}, Phys. Rev. B 93 (2016) 165416.

\bibitem{Derrien2014a}
T.~J.-Y. Derrien, R.~Koter, J.~Kr\"{u}ger, S.~H\"{o}hm, A.~Rosenfeld, J.~Bonse,
  \href{http://scitation.aip.org/content/aip/journal/jap/116/7/10.1063/1.4887808}{Plasmonic
  formation mechanism of periodic 100-nm-structures upon femtosecond laser
  irradiation of silicon in water}, J. Appl. Phys. 116~(7) (2014) 074902.

\bibitem{Mueller2010}
R.~M\"{u}ller, J.~Bethge, Plasmonic decay in a metallic grating after
  femtosecond pulse excitation, Phys. Rev. B 82 (2010) 115408.

\revadd{\bibitem{Hohenau2008}
A.~Hohenau, A.~Drezet, M.~Weissenbacher, F.~R. Aussenegg, and J.~R. Krenn,
  Effects of damping on surface-plasmon pulse propagation and
  refraction, Phys. Rev. B 78 (2008) 155405.}

\bibitem{Jackson1999}
J.~D. Jackson, Classical {E}lectrodynamics, 3rd Edition, Wiley, 1999.

\bibitem{Johnson1972}
P.~B. Johnson, R.-W. Christy, Optical constants of the noble metals, Phys. Rev.
  B 6~(12) (1972) 4370.

\bibitem{Kubo2005}
A.~Kubo, K.~Onda, H.~Petek, Z.~Sun, Y.~S. Jung, H.~K. Kim,
  \href{http://pubs.acs.org/doi/abs/10.1021/nl0506655}{Femtosecond imaging of
  surface plasmon dynamics in a nanostructured silver film}, Nano Lett. 5~(6)
  (2005) 1123--1127.

\bibitem{Ordal1985}
M.~A. Ordal, R.~J. Bell, R.~W.~A. Jr., L.~L. Long, M.~R. Querry, Optical
  properties of fourteen metals in the infrared and far infrared: {Al}, {Co},
  {Cu}, {Au}, {Fe}, {Pb}, {Mo}, {Ni}, {Pd}, {Pt}, {Ag}, {Ti}, {V}, and {W},
  Appl. Opt. 24 (1985) 4493--4499.

\bibitem{Hagemann1975}
H.-J. Hagemann, W.~Gudat, C.~Kunz, Optical constants from the far infrared to
  the x-ray region: {Mg}, {Al}, {Cu}, {Ag}, {Au}, {Bi}, {C}, and
  {A1}$_2${O}$_3$, J. Opt. Soc. Am. B 65~(6) (1975) 742--744.

\bibitem{Kedenburg2012}
S.~Kedenburg, M.~Vieweg, T.~Gissibl, H.~Giessen, Linear refractive index and
  absorption measurements of nonlinear optical liquids in the visible and
  near-infrared spectral region, Opt. Mater. Express 2~(11) (2012) 1588--1611.

\bibitem{Postava2000}
K.~Postava, H.~Sueki, M.~Aoyama, T.~Yamaguchi, C.~Ino, Y.~Igasaki, M.~Horie,
  \href{http://scitation.aip.org/content/aip/journal/jap/87/11/10.1063/1.373461}{Spectroscopic
  ellipsometry of epitaxial {ZnO} layer on sapphire substrate}, J. Appl. Phys.
  87~(11) (2000) 7820.

\bibitem{Krishnan1994}
S.~Krishnan, C.~D. Anderson, P.~C. Nordine, Optical properties of liquid and
  solid zirconium, Phys. Rev. B 49~(5) (1994) 3161.

\end{thebibliography}

\appendix

\section{\revadd{ Optical constants of materials}\label{sec:AppendixOptical}}

In this work, several sources of optical data were used. \revadd{For Al$_2$O$_3$,} Al, Au, Be, a-C, Co, Cr, Cu, diamond, Ethanol, Fe, GaAs, GaP, Ge, Graphite, H$_2$O, InP, Li, Mo, Na, Nb, Ni, Pt, c-Si, a-Si, SiC, SiO$_2$, Ta,  Ti, TiO$_2$, V, and W, the optical data were mostly taken from Ref. \cite{Palik1985}. For Mg, the work of \emph{Hagemann et al.} was used. \cite{Hagemann1975} For ZnO, the work of \emph{Postava et al.} \cite{Palik1985}. The detailed optical data used in this work are summarized in Tab. \ref{tab:OpticalProperties}.

\begin{table}
\begin{centering}
\begin{tabular}{c|cc|cc|c}
\multicolumn{1}{c}{} &  & \multicolumn{1}{c}{} &  & \multicolumn{1}{c}{} & \tabularnewline
\hline 
\hline 
\multirow{3}{*}{{\footnotesize{}Material}} & \multicolumn{4}{c|}{{\footnotesize{}Wavelength}} & \multirow{3}{*}{{\footnotesize{}Reference}}\tabularnewline
 & \multicolumn{2}{c}{{\footnotesize{}800 nm}} & \multicolumn{2}{c|}{{\footnotesize{}400 nm}} & \tabularnewline
 & {\footnotesize{}$\Re e\left(\varepsilon\right)$} & {\footnotesize{}$\Im m\left(\varepsilon\right)$} & {\footnotesize{}$\Re e\left(\varepsilon\right)$} & {\footnotesize{}$\Im m\left(\varepsilon\right)$} & \tabularnewline
\hline 
{\footnotesize{}Air} & {\footnotesize{}1.00} & {\footnotesize{}0.00} & {\footnotesize{}1.00} & {\footnotesize{}0.00} & {\footnotesize{}-}\tabularnewline
{\footnotesize{}Ag} & {\footnotesize{}-27.95} & {\footnotesize{}1.52} & {\footnotesize{}-3.77} & {\footnotesize{}0.67} & {\footnotesize{}\cite{Johnson1972}}\tabularnewline
{\footnotesize{}Al$_{2}$O$_{3}$} & {\footnotesize{}3.10 } & {\footnotesize{}0.00} & {\footnotesize{}3.19} & {\footnotesize{}0.00} & {\footnotesize{}\cite{Palik1985}}\tabularnewline
{\footnotesize{}Al} & {\footnotesize{}-63.55} & {\footnotesize{}47.31} & {\footnotesize{}-23.39} & {\footnotesize{}4.77} & {\footnotesize{}\cite{Palik1985}}\tabularnewline
{\footnotesize{}Au} & {\footnotesize{}-26.15} & {\footnotesize{}1.85} & {\footnotesize{}-1.08} & {\footnotesize{}6.49} & {\footnotesize{}\cite{Palik1985}}\tabularnewline
{\footnotesize{}Be} & {\footnotesize{}0.17} & {\footnotesize{}23.22} & {\footnotesize{}-1.42} & {\footnotesize{}18.11} & {\footnotesize{}\cite{Palik1985}}\tabularnewline
{\footnotesize{}Bi} & {\footnotesize{}-6.61} & {\footnotesize{}21.08} & {\footnotesize{}-2.32} & {\footnotesize{}6.76} & {\footnotesize{}\cite{Hagemann1975}}\tabularnewline
{\footnotesize{}a-C} & {\footnotesize{}4.80} & {\footnotesize{}3.37} & {\footnotesize{}3.53} & {\footnotesize{}3.77} & {\footnotesize{}\cite{Palik1985}}\tabularnewline
{\footnotesize{}Co} & {\footnotesize{}-17.65} & {\footnotesize{}25.20} & {\footnotesize{}-7.74} & {\footnotesize{}7.63} & {\footnotesize{}\cite{Palik1985}}\tabularnewline
{\footnotesize{}Cr} & {\footnotesize{}-1.42} & {\footnotesize{}36.29} & {\footnotesize{}-10.65} & {\footnotesize{}10.76} & {\footnotesize{}\cite{Palik1985}}\tabularnewline
{\footnotesize{}Cu} & {\footnotesize{}-25.27} & {\footnotesize{}2.52} & {\footnotesize{}-3.49} & {\footnotesize{}5.22} & {\footnotesize{}\cite{Palik1985}}\tabularnewline
{\footnotesize{}Diamond} & {\footnotesize{}5.77} & {\footnotesize{}0.00} & {\footnotesize{}6.08} & {\footnotesize{}0.00} & {\footnotesize{}\cite{Palik1985}}\tabularnewline
{\footnotesize{}Ethanol} & {\footnotesize{}1.84} & {\footnotesize{}0.00} & {\footnotesize{}-} & {\footnotesize{}-} & {\footnotesize{}\cite{Kedenburg2012}}\tabularnewline
{\footnotesize{}Fe} & {\footnotesize{}-6.38} & {\footnotesize{}20.19} & {\footnotesize{}-3.05} & {\footnotesize{}8.27} & {\footnotesize{}\cite{Palik1985}}\tabularnewline
{\footnotesize{}GaAs} & {\footnotesize{}13.55} & {\footnotesize{}0.63} & {\footnotesize{}14.53} & {\footnotesize{}18.76} & {\footnotesize{}\cite{Palik1985}}\tabularnewline
{\footnotesize{}GaP} & {\footnotesize{}10.13} & {\footnotesize{}0.00} & {\footnotesize{}17.53} & {\footnotesize{}2.31} & {\footnotesize{}\cite{Palik1985}}\tabularnewline
{\footnotesize{}Ge} & {\footnotesize{}21.96} & {\footnotesize{}3.01} & {\footnotesize{}12.24} & {\footnotesize{}18.34} & {\footnotesize{}\cite{Palik1985}}\tabularnewline
{\footnotesize{}Graphite} & {\footnotesize{}5.95} & {\footnotesize{}11.58} & {\footnotesize{}5.20} & {\footnotesize{}6.76} & {\footnotesize{}\cite{Palik1985}}\tabularnewline
{\footnotesize{}H$_{2}$O} & {\footnotesize{}1.76} & {\footnotesize{}0.00} & {\footnotesize{}1.82} & {\footnotesize{}0.00} & {\footnotesize{}\cite{Palik1985}}\tabularnewline
{\footnotesize{}InP} & {\footnotesize{}11.93} & {\footnotesize{}1.46} & {\footnotesize{}16.49} & {\footnotesize{}15.30} & {\footnotesize{}\cite{Palik1985}}\tabularnewline
{\footnotesize{}Li} & {\footnotesize{}-14.25} & {\footnotesize{}2.16} & {\footnotesize{}-2.58} & {\footnotesize{}0.96} & {\footnotesize{}\cite{Palik1985}}\tabularnewline
{\footnotesize{}Mo} & {\footnotesize{}2.08} & {\footnotesize{}24.52} & {\footnotesize{}-1.19} & {\footnotesize{}19.51} & {\footnotesize{}\cite{Palik1985}}\tabularnewline
{\footnotesize{}Mg} & {\footnotesize{}-58.35} & {\footnotesize{}11.33} & {\footnotesize{}-12.41} & {\footnotesize{}1.25} & {\footnotesize{}\cite{Hagemann1975}}\tabularnewline
{\footnotesize{}Na} & {\footnotesize{}-11.84} & {\footnotesize{}0.35} & {\footnotesize{}-2.30} & {\footnotesize{}0.21} & {\footnotesize{}\cite{Palik1985}}\tabularnewline
{\footnotesize{}Nb} & {\footnotesize{}-6.74} & {\footnotesize{}14.49} & {\footnotesize{}-0.61} & {\footnotesize{}12.90} & {\footnotesize{}\cite{Palik1985}}\tabularnewline
{\footnotesize{}Ni} & {\footnotesize{}-13.04} & {\footnotesize{}21.73} & {\footnotesize{}-2.98} & {\footnotesize{}7.60} & {\footnotesize{}\cite{Palik1985}}\tabularnewline
{\footnotesize{}Pt} & {\footnotesize{}-16.47} & {\footnotesize{}28.11} & {\footnotesize{}-5.11} & {\footnotesize{}9.77} & {\footnotesize{}\cite{Palik1985}}\tabularnewline
{\footnotesize{}a-Si} & {\footnotesize{}15.17} & {\footnotesize{}0.85} & {\footnotesize{}13.39} & {\footnotesize{}18.84} & {\footnotesize{}\cite{Palik1985}}\tabularnewline
{\footnotesize{}Si} & {\footnotesize{}13.63} & {\footnotesize{}0.048} & {\footnotesize{}30.85} & {\footnotesize{}4.30} & {\footnotesize{}\cite{Palik1985}}\tabularnewline
{\footnotesize{}SiC} & {\footnotesize{}6.78} & {\footnotesize{}0.00} & {\footnotesize{}7.64} & {\footnotesize{}0.00} & {\footnotesize{}\cite{Palik1985}}\tabularnewline
{\footnotesize{}SiO$_{2}$} & {\footnotesize{}2.11} & {\footnotesize{}0.00} & {\footnotesize{}2.16} & {\footnotesize{}0.00} & {\footnotesize{}\cite{Palik1985}}\tabularnewline
{\footnotesize{}Ta} & {\footnotesize{}-11.17} & {\footnotesize{}7.90} & {\footnotesize{}2.49} & {\footnotesize{}12.60} & {\footnotesize{}\cite{Palik1985}}\tabularnewline
{\footnotesize{}Ti} & {\footnotesize{}-6.21} & {\footnotesize{}25.2} & {\footnotesize{}-4.36} & {\footnotesize{}12.36} & {\footnotesize{}\cite{Johnson1972}}\tabularnewline
{\footnotesize{}TiO$_{2}$} & {\footnotesize{}7.78} & {\footnotesize{}0.00} & {\footnotesize{}11.55} & {\footnotesize{}0.00} & {\footnotesize{}\cite{Palik1985}}\tabularnewline
{\footnotesize{}V} & {\footnotesize{}1.88} & {\footnotesize{}21.76} & {\footnotesize{}-4.92} & {\footnotesize{}17.24} & {\footnotesize{}\cite{Palik1985}}\tabularnewline
{\footnotesize{}W} & {\footnotesize{}5.22} & {\footnotesize{}19.44} & {\footnotesize{}5.68} & {\footnotesize{}16.34} & {\footnotesize{}\cite{Palik1985}}\tabularnewline
{\footnotesize{}ZnO} & {\footnotesize{}3.80} & {\footnotesize{}0.00} & {\footnotesize{}5.11} & {\footnotesize{}0.07} & {\footnotesize{}\cite{Postava2000}}\tabularnewline
{\footnotesize{}Zr} & {\footnotesize{}-10.56} & {\footnotesize{}16.38} & {\footnotesize{}-2.81} & {\footnotesize{}7.18} & {\footnotesize{}\cite{Krishnan1994}}\tabularnewline
\hline 
\hline 
\multicolumn{1}{c}{} &  & \multicolumn{1}{c}{} &  & \multicolumn{1}{c}{} & \tabularnewline
\end{tabular}
\par\end{centering}

\caption{\label{tab:OpticalProperties}Summary of the dielectric permittivities used in this work.}

\end{table}

\end{document}